\documentclass{ws-ijmpd}
\usepackage[super,compress]{cite}
\usepackage{graphicx}
\usepackage{epstopdf}
\usepackage{tensor}
\usepackage{undertilde}

\begin{document}

\markboth{J. P. Morais Gra\c{c}a, Gonodou I. Salako and V. B. Bezerra}
{Quasinormal modes of a black hole with a cloud of strings in Einstein-Gauss-Bonnet gravity}

%
\catchline{}{}{}{}{}
%

\title{Quasinormal modes of a black hole with a cloud of strings in Einstein-Gauss-Bonnet gravity}

\author{J. P. Morais Gra\c{c}a}

\address{Universidade Federal do Paran\'a\\
Instituto de F\'isica, Curitiba, Brasil\\
jpmorais@gmail.com}

\author{Godonou I. Salako}

\address{Institut de math\'ematiques et de Sciences Physiques (IMSP)\\
 01 BP 613 Porto-Novo, B\'enin\\
inessalako@gmail.com}

\author{Valdir B. Bezerra}

\address{Universidade Federal da Para\'iba\\
 Jo\~ao Pessoa, Brasil\\
valdir@fisica.ufpb.br}

\maketitle

\begin{history}
\received{Day Month Year}
\revised{Day Month Year}
\end{history}

\begin{abstract}
The quasinormal modes for a scalar field in the background
spacetime corresponding to a black hole, with a cloud of strings, in Einstein-Gauss-Bonnet gravity, and the tensor quasinormal modes corresponding to perturbations in such spacetime, were both calculated using the WKB approximation. In the obtained results we emphasize
the role played by the parameter associated with the string cloud, comparing them with the results already obtained for the Boulware-Deser metric. We also study how the Gauss-Bonnet correction to general relativity affects the results for the quasinormal modes, comparing them with the same background in general relativity.
\end{abstract}

\keywords{Quasinormal modes, modified gravity, cloud of strings}



\section{Introduction}
\label{intro}

String theory is a promising framework for a quantum theory of gravity. Its main idea concerns on the fact that the building blocks of nature are one-dimensional strings, instead of zero dimensional particles. Due to inflation, these fundamental strings could have been stretched in the early universe and can exist today, at least hypothetically, as extended strings of cosmological sizes \cite{Copeland}. A more traditional framework to describe particles and its interactions, the quantum theory of fields, also predicts that cosmic strings could have been formed in the early universe due to a phase transition \cite{Hindmarsh:1994re}. For these reasons, it is natural to study the cosmological and astrophysical effects generated by one-dimensional extended strings, as individual objects or also as a collective phenomena.

A cloud of strings is the one-dimensional analogous of a cloud of dust. The theoretical framework for such objects has been developed by Letelier in \cite{Letelier:1979ej}, and recently several works on the subject has appeared, e.g., on its space-time structure \cite{Ghosh:2014dqa,Ghosh:2014pga,Mazharimousavi:2015sfo}, and thermodynamics properties \cite{Herscovich:2010vr}. In the spirit of considering these strings as fundamental objects, it is natural to consider them in a gravitational theory that goes beyond Einstein's gravity, since the low-energy limit of string theory implies the natural generalization of general relativity to higher dimensions, namely, Lovelock theory of gravity. 

In this paper, our interest is to study the perturbations around the classical space-time generated by the cloud of strings. To be more precise, we will study the scalar and tensor quasinormal modes for a cloud of strings inside a black hole. Quasinormal modes are complex numbers that model the emission of gravitational waves by perturbed objects. Its real part is related with the frequency of emission, and its imaginary part is related with how the oscillation decay in time (For a review, see \cite{Berti:2009kk}).

The study of quasinormal modes has received a great attention in the last years for two fundamental reasons: The first and most obvious one is related with the search for gravitational waves emitted by compact objects, such as black holes and neutron stars. The recent first direct detection of an emission, and the amazingly precise matching between the observed and theoretically calculated data, has crowned such efforts \cite{Abbott:2016blz}. As a secondary reason, we can mention the gauge/gravity dualities, where the quasinormal modes in the gravitational side of the correspondence is related with the poles of a propagator in the dual theory. Being a mere theoretical framework for the study of strong coupled gauge theories, or a realistic description for a hypothetical holographic universe, the fact is that the gauge/gravity dualities has attained a great deal of attention over the last years.

Our main goal in this paper will be to study the quasinormal modes for a black hole, with a cloud of strings, in the 2nd order Lovelock theory, also called Einstein-Gauss-Bonnet gravity. The limit to general relativity can be easily obtaining taking $\alpha \rightarrow 0$, where $\alpha$ is the parameter that takes care of the difference between general relativity and Einstein-Gauss-Bonnet gravity. The quasinormal modes for a black hole in Einstein-Gauss-Bonnet has already been calculated in \cite{Konoplya:2004xx} and \cite{Chakrabarti:2006ei}, for scalar and tensor modes, respectively, and its stability criteria has been studied since then \cite{Cuyubamba:2016cug,Konoplya:2008ix,Abdalla:2005hu}.  Now we add a cloud of strings to understand how the presence of this object can affect the frequency of emission and the rate of decay. Also, we will be able to study how the introduction of the Gauss-Bonnet term will modify the spectra of emission generated by the cloud of strings, inside a black hole, as compared with the same system in general relativity. 

Gauss-Bonnet gravity can be considered for dimension greater or equal than 5, but we will not consider dimensions 5 or 6. The former because there is no solution for the whole range of the Gauss-Bonnet parameter, and the latter since it is unstable to tensor mode perturbations \cite{Dotti:2004sh}. We have chosen to work with dimensions 7 and 9 to compare the role of the dimensionality of the space-time on the behaviour of the quasinormal modes. In this paper we will not consider phenomenological bounds as an upper limit in the Gauss-Bonnet parameter. Despite the fact that it is of fundamental importance to constraint the parameters of the theory with observations, we will keep the parameters free of bounds, with the exception of the already mentioned instability for dimension d=6. For a discussion on parameter bounds for some modified theories of gravity, see \cite{Psaltis:2008bb,Will,Calmet:2008tn,Atkins:2012yn,Xianyu:2013rya,Onofrio:2010zz,Onofrio:2014txa,Wegner:2015aea}.

This paper is organized in the following manner: In section 2 we will briefly discuss how to construct the Lovelock model of gravity. In section 3 we will introduce the formalism for the cloud of strings, together with their space-time structure. In section 4 we will calculate the scalar quasinormal modes for such a system and compare then with the same gravitational model, but without the cloud of strings, with and without the Gauss-Bonnet additional term. In section 5 we will do the same calculation, but now for the tensor quasinormal modes. It will be interesting to note that, as the energy of the cloud of strings grows, the scalar and tensor spectra approach each other, achieving some kind of isospectrality. Finally, in section 6 we will present our conclusions.
  
\section{Lovelock models of gravity}
\label{sec:1}

Lovelock models of gravity are a natural generalization of general relativity to higher dimensions. Their field equations are up to second order in derivatives of the metric, and constructed in such a way that the energy-momentum tensor of the theory is conserved. In terms of a local frame $e_{(a)} = e\indices{_{(a)}^\mu} \partial_\mu$, such that  $e\indices{_{(a)}^\mu} e\indices{_{(b)}^\nu} g_{\mu\nu} = \eta_{(a)(b)}$, where $\eta_{(a)(b)}$ is the Minkowski metric, the n-th order Lovelock model of gravity is given by

\begin{equation}
S = S_{0} + S_{1} + ... + S_{n},
\end{equation}
where 

\begin{eqnarray}
\nonumber
S_{n} =&& \frac{\alpha_n}{(d-2n)!} \int \epsilon_{(a_1)(b_1)...(a_n)(b_n)(c_1)...(c_{d-2n})} e^{(c_1)} \wedge ... 
\\
&&...\wedge e^{(c_{d-2n})} \wedge \rho^{(a_1)(b_1)} \wedge ... \wedge \rho^{(a_n)(b_n)},
\end{eqnarray}
with $\rho^{(a)(b)}$ being a 2-form connection, and $d$ the spacetime dimension. The zero-order action is a cosmological constant term. The first-order is the Einstein-Hilbert action. The second-order term is called Gauss-Bonnet term. As an example, in $5$ dimensions it is given by

\begin{equation}
S_{2} = \alpha_2 \int \epsilon_{(a)(b)(c)(d)(c_1)} e^{(c_1)} \wedge \rho^{(a)(b)} \wedge \rho^{(c)(d)}.
\label{ch1:eqGBformas}
\end{equation}

The $2$-form of curvature can be obtained from the second Cartan structure equation, $\rho^{(a)(b)} = d\omega^{(a)(b)} + \omega\indices{^{(a)}_{(c)}} \wedge \omega^{(c)(b)}$. By its turn, $\omega^{(a)(b)}$, called spin connection, can be obtained in terms of the \textit{vielbein}, given by $e^{(a)} e_{(b)} = \delta^{(a)}_{(b)}$, using the first Cartan structure equation, $ d e^{(a)} + \omega\indices{^{(a)}_{(b)}} \wedge e^{(b)} = 0$, where the spacetime is torsion-free.

In coordinate indices, the action of the second-oder Lovelock term reduces to

\begin{equation}
S_{2} = - \alpha_2 \int d^dx \sqrt{-g} (R^2 + R_{\mu\nu\alpha\beta} R^{\mu\nu\alpha\beta} - 4R^{\mu\nu}R_{\mu\nu}) ,
\end{equation}
where $g$ is the determinant of the metric $g_{\mu\nu}$. In this paper we will consider both the first and second terms of the action. This theory is usually called Einstein-Gauss-Bonnet gravity. The field equations can be obtained, as usual, by varying the action with respect to the metric. In coordinate indices, it is given by

\begin{equation}
\mathcal{G}_{\mu\nu} = \kappa^{-1}(G_1)_{\mu\nu} + \alpha_2 (G_2)_{\mu\nu} = T_{\mu\nu},
\label{EGBequation}
\end{equation}
where $\kappa^{-1} = 8 \pi G$, but we will set it equal to unity, to be consistent with \cite{Dotti:2005sq} \footnote{We should not that in \cite{Chakrabarti:2006ei}, the authors used $G=1$, and not $\kappa=1$.}. The tensor $(G_1)_{\mu\nu}$ is the Einstein tensor, given by $(G_1)_{\mu\nu} = R_{\mu\nu} - (1/2)g_{\mu\nu} R$. By its turn, $(G_2)_{\mu\nu}$ is given by

\begin{eqnarray}
(G_2)_{\mu\nu} &=& 2 R\indices{_\mu_{\alpha_1\alpha_2\alpha_3}}R\indices{_\nu^{\alpha_1\alpha_2\alpha_3}} - 4 R_{\mu\alpha_1 \nu\alpha_2} R^{\alpha_1\alpha_2} - 4 R_{\mu\alpha_1} R\indices{^{\alpha_1}_\nu} + 2 R R_{\mu\nu} 
\nonumber
\\
&-& \frac{1}{2} g_{\mu\nu}\left(R^2 - 4 R_{\alpha_1\alpha_2}R^{\alpha_1\alpha_2} + R_{\alpha_1\alpha_2\alpha_3\alpha_4}R^{\alpha_1\alpha_2\alpha_3\alpha_4}\right).
\end{eqnarray}

Let us consider a Schwarzschild-like metric with spherical symmetric, 

\begin{equation}
ds^2 = - f(r) dt^2 + \frac{1}{f(r)} dr^2 + r^2 d\Omega,
\label{metric}
\end{equation}
where $d\Omega$ is the ($d-2$)-sphere in a $d$ dimensional spacetime. A vacuum solution of the Einstein-Gauss-Bonnet field equations (\ref{EGBequation}) is the Boulware-Deser metric,

\begin{equation}
f(r) = 1 + \frac{r^2}{2 \alpha} \left(1 \pm \sqrt{1 - \frac{C_1 8 \alpha}{(d-2) r^{d-1}}} \right),
\label{ch5:solucaoBD}
\end{equation}
where $\alpha = \alpha_2 (d-4)(d-3)$ and $C_1$ is an integral constant usually chosen so that the Boulware-Deser metric reduces to Schwarzschild metric as $\alpha_2 \rightarrow 0$. In this paper we are interest to solve the Einstein-Gauss-Bonnet equations with a energy-momentum tensor due to a string cloud.

\section{The cloud of strings}

In cosmology, it is usual to consider clouds of matter to model the cold content of the universe. A cloud of strings is an analogue model, where the zero-dimensional content is replaced by one-dimensional objects, extended along some defined direction. In the present paper, we will consider strings distributed in a spherical symmetric pattern.

In the same way as a particle generates a word-line in spacetime, a string generate a two-dimensional world-sheet $\Sigma$, parametrized by the parameters $\lambda^0$ and $\lambda^1$. The action for a single string is given by the Nambu-Goto action, 

\begin{equation}
S_{GN} = m \int_{\Sigma} \sqrt{-\gamma} d\lambda^0 d\lambda^1, 
\label{NGaction}
\end{equation}
where $m$ is positive and related to the tension of the string, and $\gamma$ is the determinant of the induced metric 

\begin{equation}
\gamma_{ab} = g_{\mu\nu} \frac{\partial x^\mu}{\partial \lambda^a} \frac{\partial x^\nu}{\partial \lambda^b}.
\end{equation}

The action can also be described by a spacetime bi-vector $\Sigma^{\mu\nu}$, given by

\begin{equation}
\Sigma^{\mu\nu} = \epsilon^{ab} \frac{\partial x^\mu}{\partial \lambda^a} \frac{\partial x^\nu}{\partial \lambda^b},
\end{equation}
such that the Nambu-Goto action can be written as

\begin{equation}
S_{GN} = m \int_{\Sigma} \sqrt{-\frac{1}{2} \Sigma_{\mu\nu} \Sigma^{\mu\nu}} d\lambda^0 d\lambda^1. 
\label{NGaction2}
\end{equation}
The action (\ref{NGaction2}) allow us to calculate the energy-momentum tensor for the string in a straightforward way, from the relation $T_{\mu\nu} = -2 \partial \mathcal{L} / \partial g^{\mu\nu}$. The energy-tensor of a single string is then given by

\begin{equation}
T^{\mu\nu} = m \frac{\Sigma^{\mu\sigma} \Sigma\indices{_\sigma^\nu}}{\sqrt{-\gamma}},
\end{equation}
and a string cloud can be naturally described by a density of strings, such as $T_{cloud}^{\mu\nu} = \rho T^{\mu\nu}$, where $\rho$ is the density of the string cloud.

For a spherically symmetric string configuration, the only non-null component of the bi-vector is $\Sigma^{tr} = - \Sigma^{rt}$. Then, the only non-null components of the energy-momentum tensor of a cloud of strings are given by

\begin{equation}
T\indices{^t_t} = T\indices{^r_t} = -\rho m \Sigma\indices{^t_r}.
\end{equation}

The conservation of the energy-momentum tensor, $\nabla_\mu T^{\mu\nu}$, allow us to calculate the radial dependence of it. The equation $\partial_\mu (T\indices{^t_t} \sqrt{-g})$ = 0, have as solution

\begin{equation}
T\indices{^t_t} = - \frac{\eta^2}{r^{d-2}},
\end{equation}
where $\eta^2$ is a integration constant that is related with the energy-density of the whole cloud, and depends both on the energy of individual strings and on the string cloud density. The energy-momentum tensor of a cloud of strings resembles the energy of a global monopole \cite{Barriola:1989hx}, since both describes radial energy flux.

The line element (\ref{metric}) indicates that only one function is necessary to describe the metric, namely, $f(r)$. As $T^i_i = 0 $, $i$ representing any indice of the $(d-2)$-sphere, the function $f(r)$ should obey the vacuum equations for the spherical indices. Thus, we have

\begin{equation}
\mathcal{G}\indices{^i_i} = 0,
\label{eqSphericalLovelock}
\end{equation}
where the indice $i$ is not contracted. In general relativity (first-order Lovelock gravity), the solution to (\ref{eqSphericalLovelock}) is given by

\begin{equation}
f(r) = 1 + \frac{C_1}{r^{d-3}} + \frac{C_2}{r^{d-4}},
\end{equation}
where $C_1$ and $C_2$ are integration constants. The constant $C_1$ can be found using the Newtonian limit from general relativity, and the constant $C_2$ will be fixed by the other field equations. We can note that in $d=4$, the cloud of strings (and the global monopole) produces an angular deficit. 

We are interested, however, to solve the Einstein-Gauss-Bonnet model. The easiest way to solve the field equations is to determine the vacuum $\mathcal{G}\indices{^i_i}$ components, and to fix the integration constant inserting the obtained function in the $(t,t)$ component of the field equations. This gives us an answer which can be written as

\begin{equation}
f(r) = 1 + \frac{r^2}{2 \alpha} \left(1 \pm \sqrt{1 + \frac{8 M\alpha}{r^{d-1}} + \frac{8 \eta^2 \alpha}{(d-2)r^{d-2}}} \right),
\end{equation}
where the mass $M$ term was fixed by demanding that the above solution reduces to the Schwarzschild-Tangherlini metric as $\eta^2 = 0$ and $\alpha \rightarrow 0$. 

\section{Scalar perturbations}

\begin{figure}[]
\centering
\begin{tabular}{@{}cc@{}}
\includegraphics[scale=0.35]{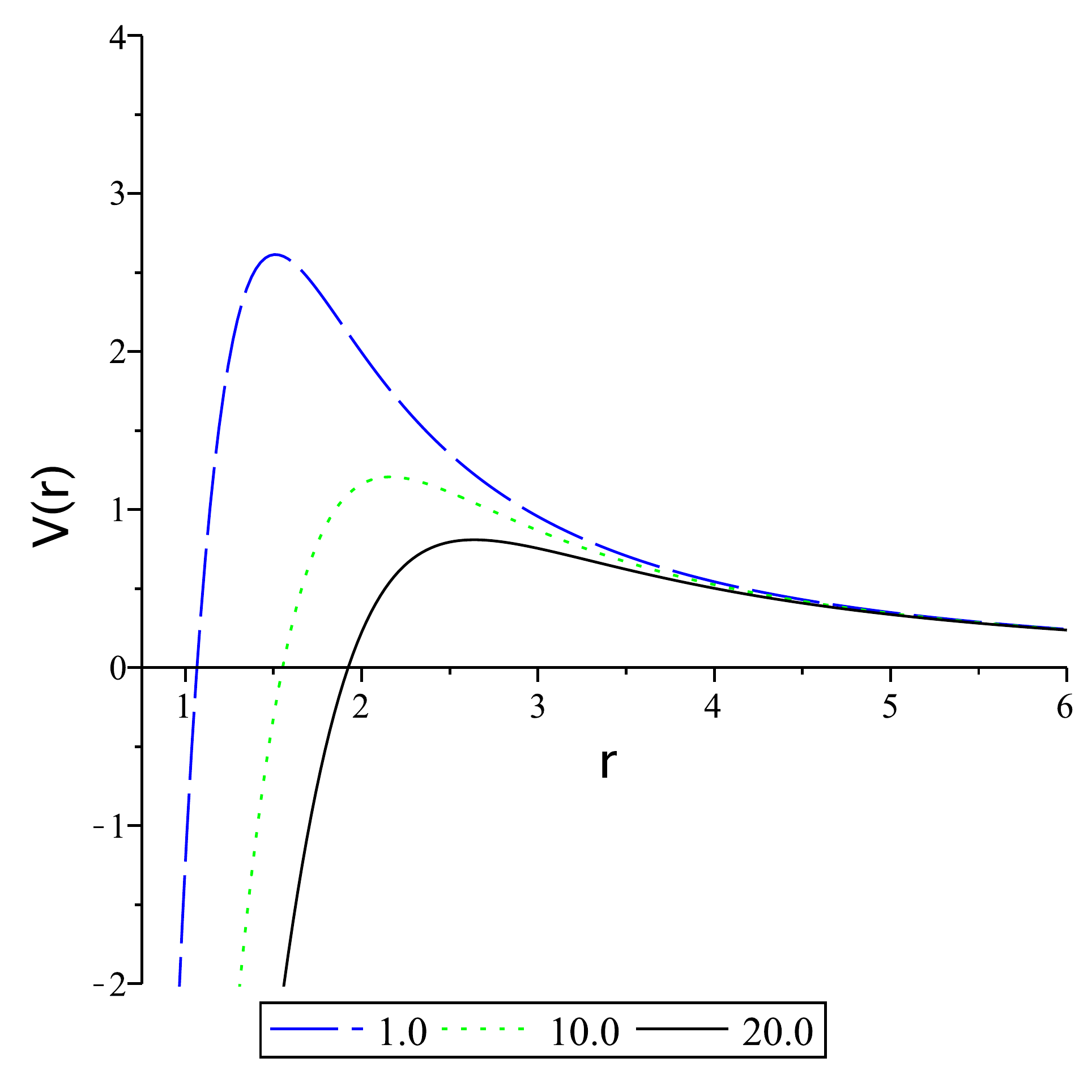} &
\includegraphics[scale=0.35]{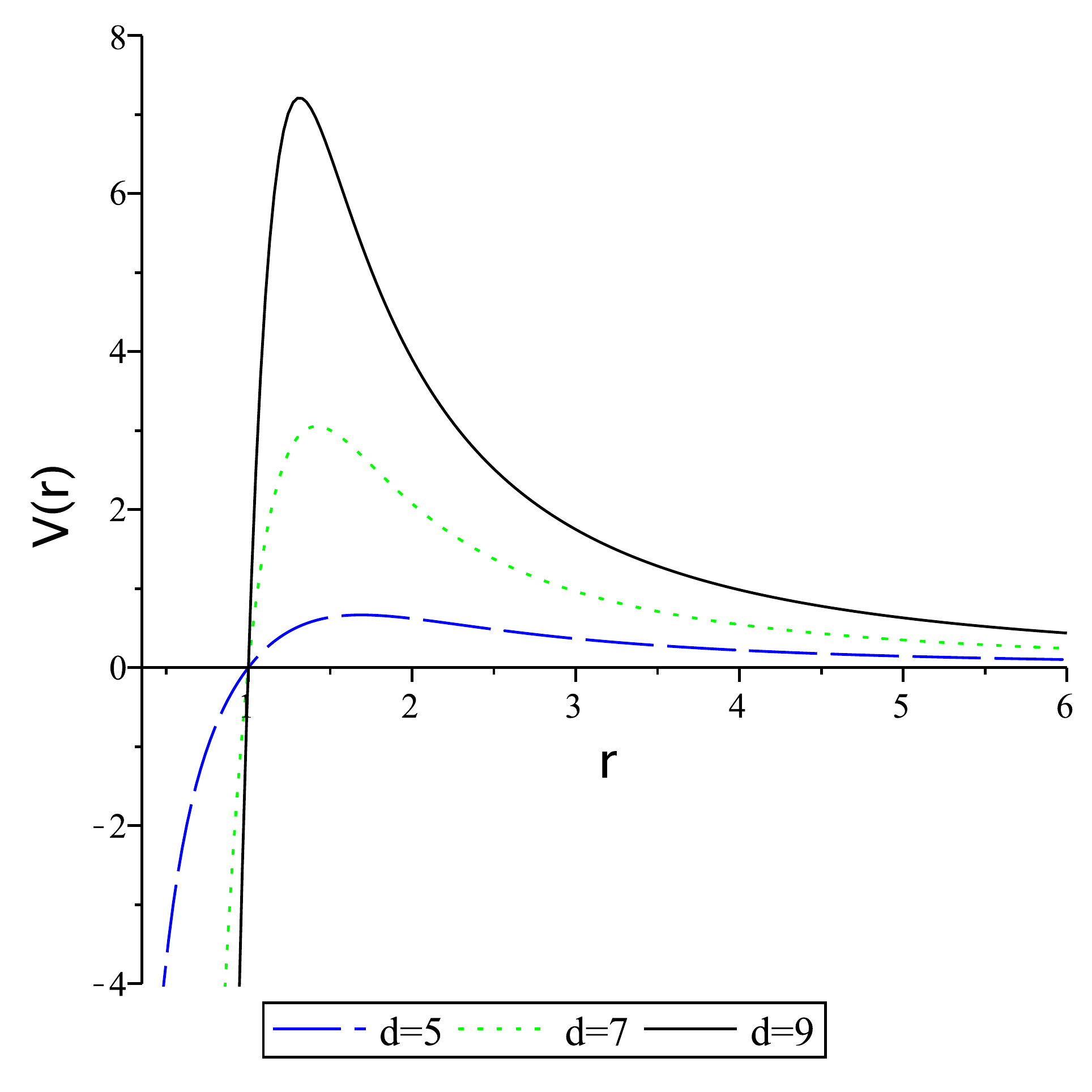} \\
\includegraphics[scale=0.35]{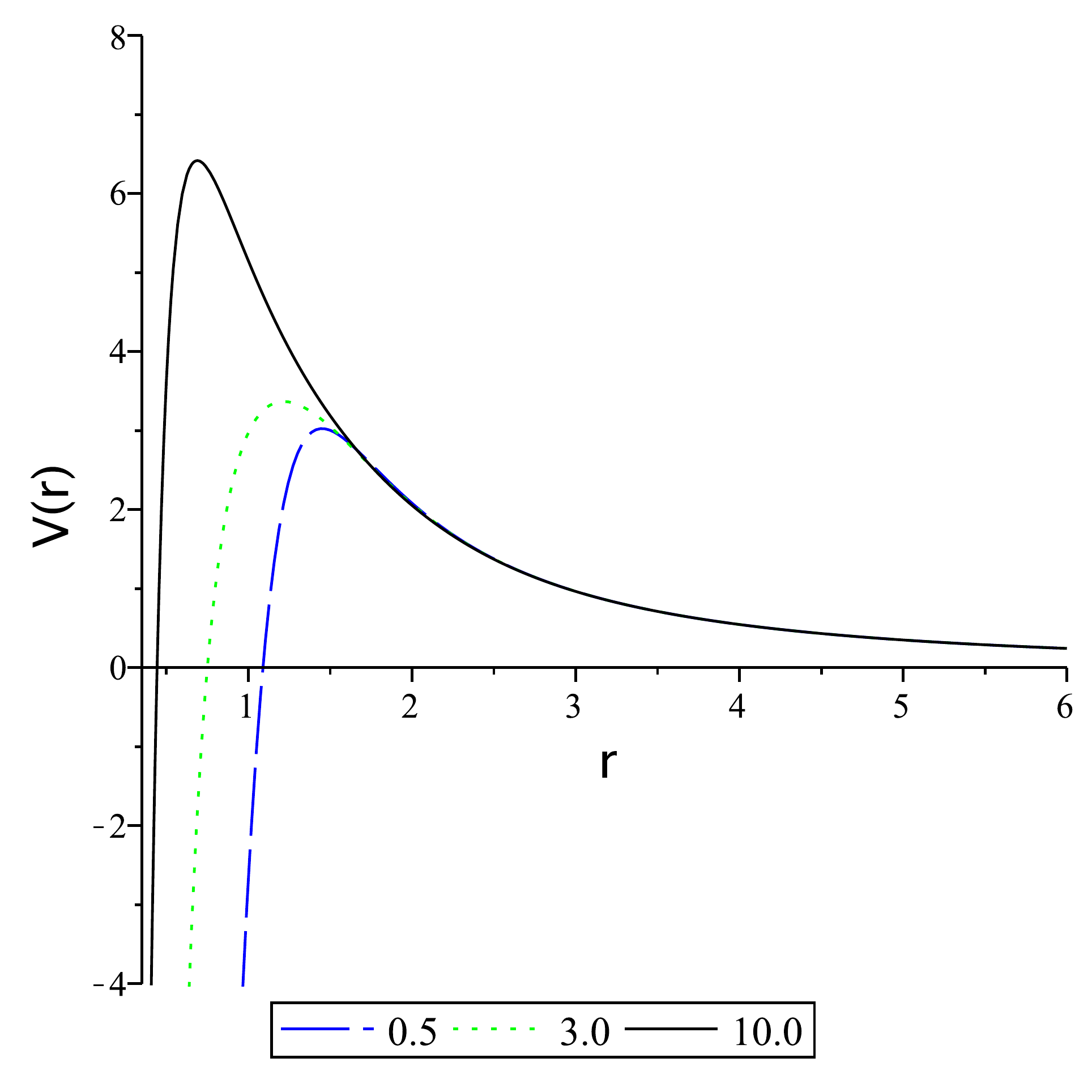} &
\includegraphics[scale=0.35]{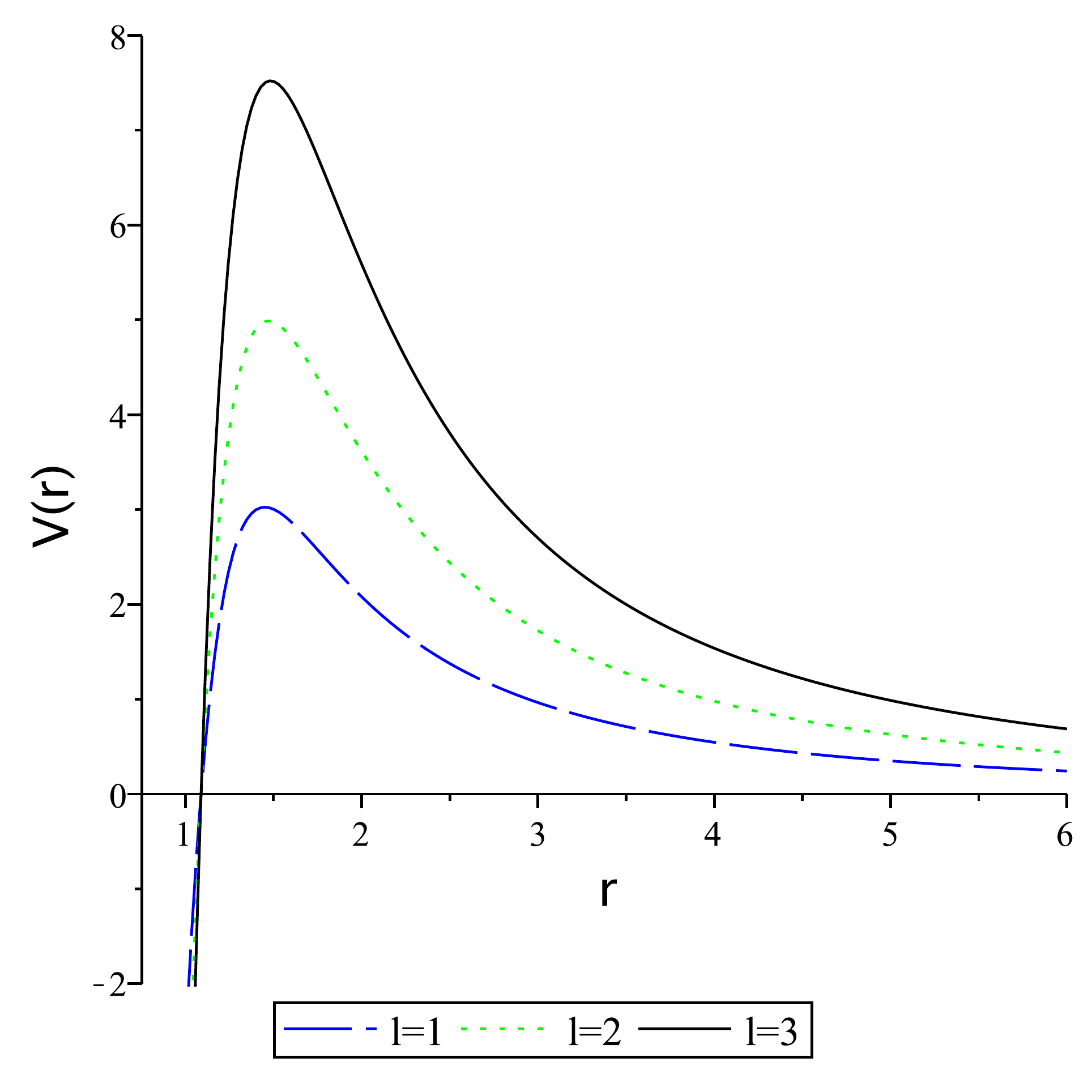} \\
\end{tabular}
\caption{Effective potential for a black hole, with a cloud of strings, in Einstein-Gauss-Bonnet gravity. Top Left: $\alpha=1, l=1, d=7$ for $\eta^2=1, 10$ and $20$. Top Right: $\alpha=1, \eta^2=0, l=1$ for $d=5,7$ and $9$. Bottom left: $\eta^2=0, l=1,d=7$ for $\alpha=0.5,3$ and $10$. Bottom right: $\alpha =0.5, \eta^2 = 0, d=7$ for $l=1,2$ and $3$. }
\label{potential}
\end{figure}

The dynamics of a massless scalar field $\Phi(\textbf{x})$, in a $d$ dimensional spacetime, obeys the Klein-Gordon equation, $\nabla_\mu \nabla^\mu \Phi(\textbf{x}) = 0$. Since we are considering a static background, the field equation can be separated as $\Phi(\textbf{x}) = e^{i\omega t} Y(\theta,\phi)^l_m \varphi(r)/r^{d-3}$, where $Y^l_m$ are the usual spherical harmonics. Then, the Klein-Gordon equation can be reduced to a Schr\"odinger type equation, namely,

\begin{equation}
\left( \frac{d^2}{dr^{*2}} + \omega^2 - V(r) \right) \Phi(\textbf{r}) = 0,
\label{scalarEquation}
\end{equation}
where the tortoise coordinate is defined as $dr/dr^* = f(r)$,  and the effective potential $V(r)$ is given by

\begin{equation}
V(r) = f(r) \left(\frac{(d-2)(d-4)}{4r^2}f(r) + \frac{d-2}{2r}f'(r) + \frac{l(l+D-3)}{r^2}\right),
\end{equation}
with $l$ being the angular momentum eigenvalue related with the angular momentum operator $L^2$. In figure (\ref{potential}) it is plotted several configuration for the potential, varying the parameters $\alpha$ and $\eta^2$, the angular momentum $l$ and the dimension $d$. We can note that, in all of the situations, the potential appears to be positive in the limit $r \rightarrow \infty$. On the top and bottom left plots, where the parameters $\eta^2$ and $\alpha$ are varying, the peak and height of the potential changes, but asymptotically it looks like the same. 

To find the scalar quasinormal modes for the scalar field, we must solve equation (\ref{scalarEquation}), together with the following set of boundary conditions,

\begin{equation}
\Phi(\textbf{x} \rightarrow - \infty) = C_I \Phi_I^{out}, \hspace{10pt} \Phi(\textbf{x} \rightarrow + \infty) = C_{II} \Phi_{II}^{out}.
\label{boundaryConditions}
\end{equation}

Regions $I$ and $II$ are away from the peak of the potential in opposite directions, and the boundary conditions (\ref{boundaryConditions}) states that we should consider the absence of incoming waves. 

Equation (\ref{scalarEquation}) can be solved using the familiar WKB approach, where an approximate WKB solution is found at both regions $I$ and $II$, and matched with the approximated solution found around the peak of the potential.  To reconcile the boundary conditions (\ref{boundaryConditions}) with the matching, the modes should obey the constraint

\begin{equation}
i\frac{Q_0}{\sqrt{2 Q_0''}} - \sum_{k=2}^{m} \Lambda_k = n + 1/2,
\end{equation}
where $Q(r) = \omega^2 - V(r)$, and $Q_0'' = d^2Q(r) / dr^{*2}$. The parameter $n$ is called the $tone$ number of the quasinormal mode.  The $\Lambda_k$'s can be obtained from the potential $Q(r)$ up to its $2k$-derivative, and indicates the order of the WKB method. The 3th order WKB was developed in \cite{Iyer:1986np} and the 6th order in \cite{Konoplya:2003ii}. The explicit formula for the considered $\Lambda_k$'s can be found in the original papers \cite{Iyer:1986np}\cite{Konoplya:2003ii}. For $l > n$, the WKB method gives us a good approximation for the correct quasinormal modes, and we will use the $5-th$ order WKB method to calculate the scalar quasinormal modes.

\begin{table}[]
\tbl{Scalar quasinormal modes for $d=7$, $l=2$ and $n=0$.}
{
\begin{tabular}{p{0.2\textwidth}p{0.2\textwidth}p{0.2\textwidth}p{0.2\textwidth}}
\hline
 $\alpha$  & $\eta^2$  & Re($\omega$) & -Im($\omega$)  \\ \hline
 0.5  & 0.1  & 2.138 & 0.496  \\
 0.5  & 1  & 1.998 & 0.453  \\
 0.5  & 10  & 1.369 & 0.292  \\
 0.5  & 20  & 1.121 & 0.239  \\
 0.5  & 30  & 0.989 & 0.211  \\
 5.0  & 0.1  & 2.647 & 0.328  \\
 5.0  & 1  & 2.474 & 0.300  \\
 5.0  & 10  & 1.572 & 0.200  \\
 5.0  & 20  & 1.221 & 0.170  \\
 5.0  & 30  & 1.052 & 0.160  \\
 10.0  & 0.1  & 3.392 & 0.430  \\
 10.0  & 1  & 3.200 & 0.398  \\
 10.0  & 10  & 1.990 & 0.212  \\
 10.0  & 20  & 1.438 & 0.140  \\ 
 10.0  & 30  & 1.187 & 0.128  \\ 
\end{tabular}
\label{tab:QNM_scalar}}
\end{table}

\begin{figure}[]
\centering
\begin{tabular}{@{}cc@{}}
\includegraphics[scale=0.75]{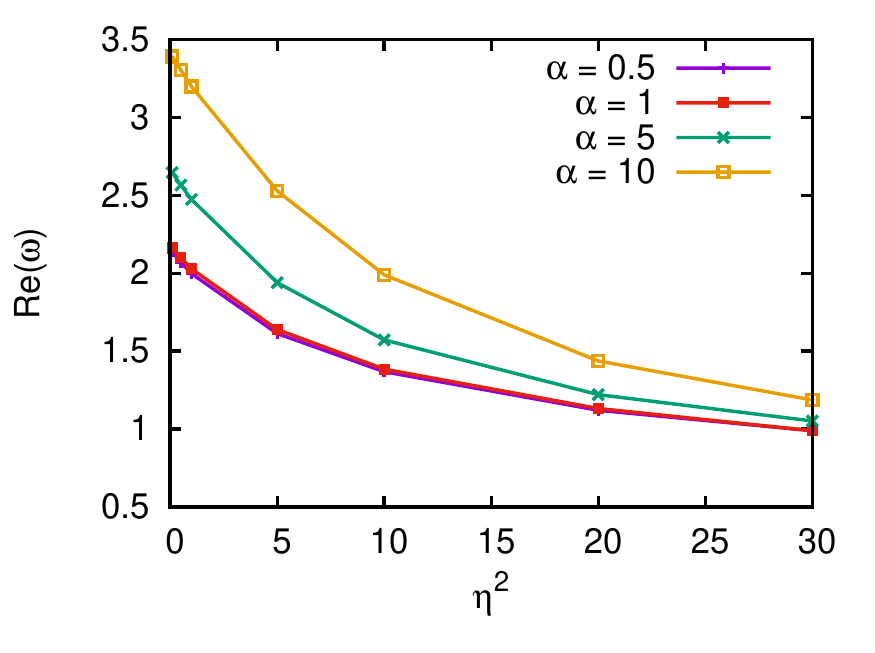} &
\includegraphics[scale=0.75]{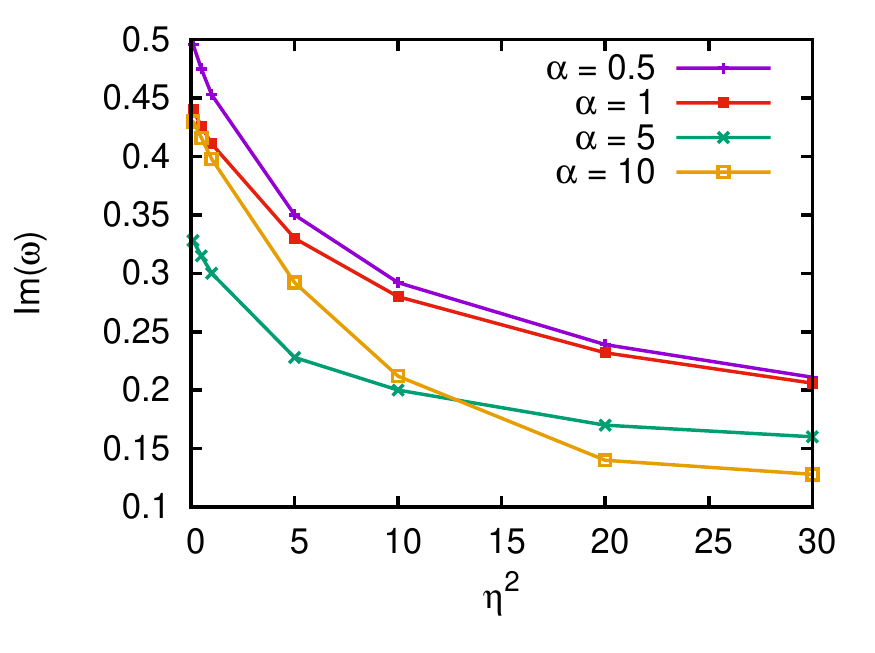} \\
\includegraphics[scale=0.75]{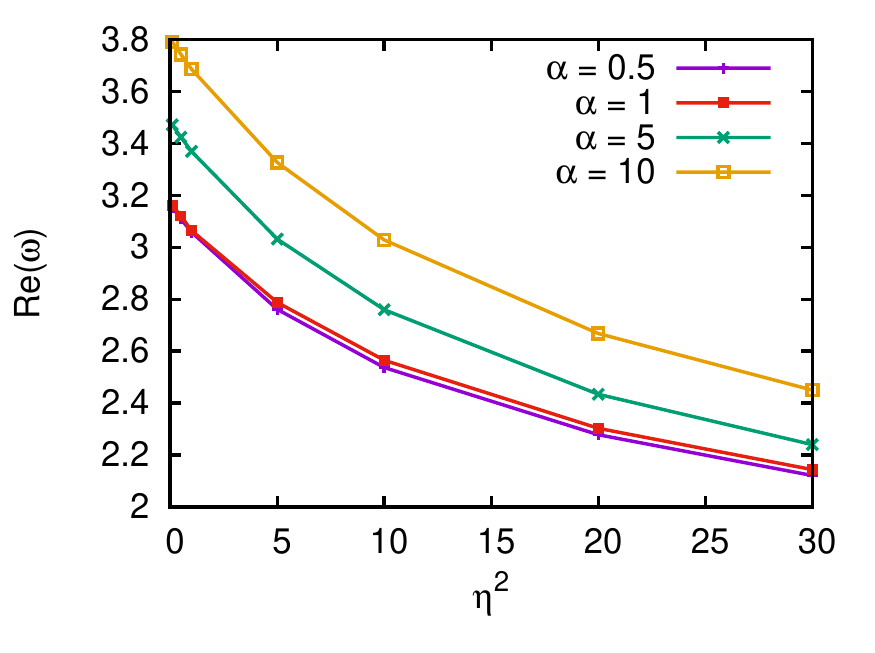} &
\includegraphics[scale=0.75]{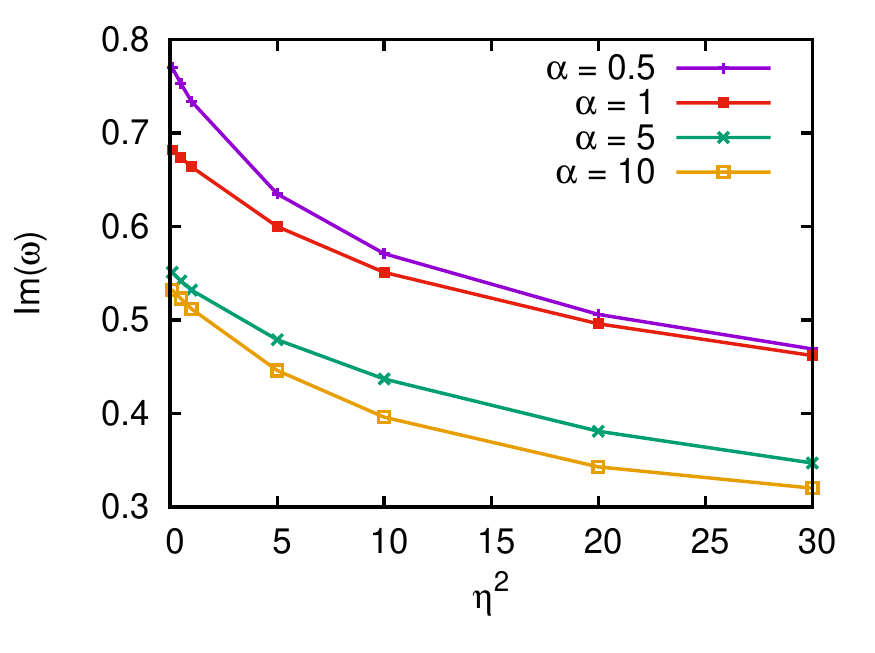} \\
\end{tabular}
\caption{Scalar quasinormal modes for a black hole, with a cloud of strings, in Einstein-Gauss-Bonnet gravity. The parameters are $l=2$ and $n=0$, for $d=7$ (top) and $d=9$ (bottom). We vary both $\alpha$ and $\eta^2$.}
\label{EGBvariaEta}
\end{figure}

The calculated values for some set of parameters is listed in table (\ref{tab:QNM_scalar}) for $d=7$. The behaviour of these values for the quasinormal modes, as compared with the quasinormal modes for the Boulware-Deser metric, are best understood in a plot. In figure (\ref{EGBvariaEta}) we plot four graphs, where we show how the quasinormal modes changes as we vary the parameter $\eta^2$, related with the cloud of strings. The top plots were calculated for $d=7$, and the bottom plots for $d=9$. We can see that both the real and imaginary values decrease as the parameter $\eta^2$ increases, indicating that the main effect of the cloud of strings is to lower the frequency and attenuate the decay. For the real part, the pattern on how the frequency decreases appear to be independent of the parameter $\alpha$, at least for small values of the parameter $\eta^2$. For the imaginary part, we cannot find a well-defined pattern, but the effect of the cloud of strings on the attenuation of the decay is clear.

\begin{figure}[]
\centering
\begin{tabular}{@{}cc@{}}
\includegraphics[scale=0.75]{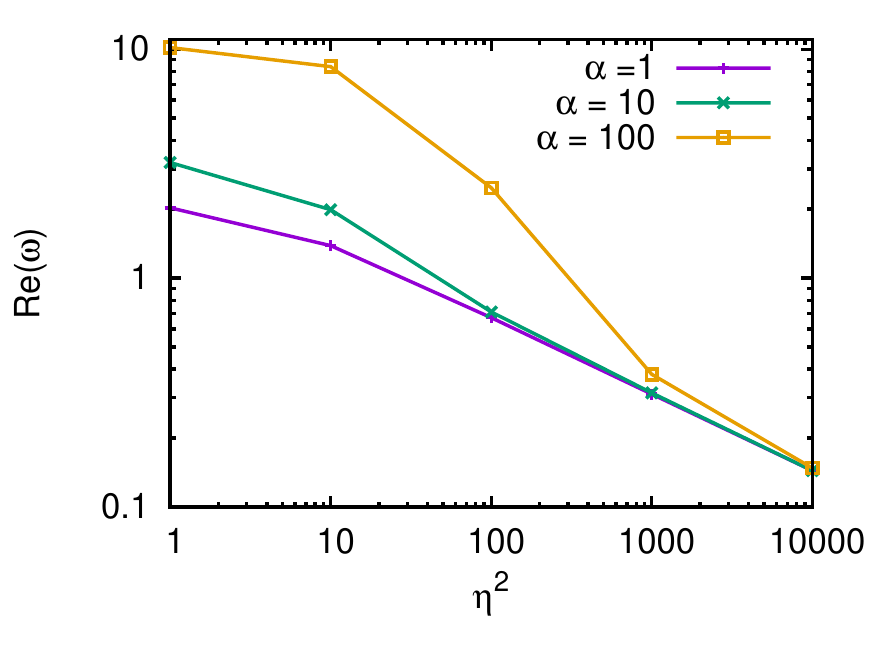} &
\includegraphics[scale=0.75]{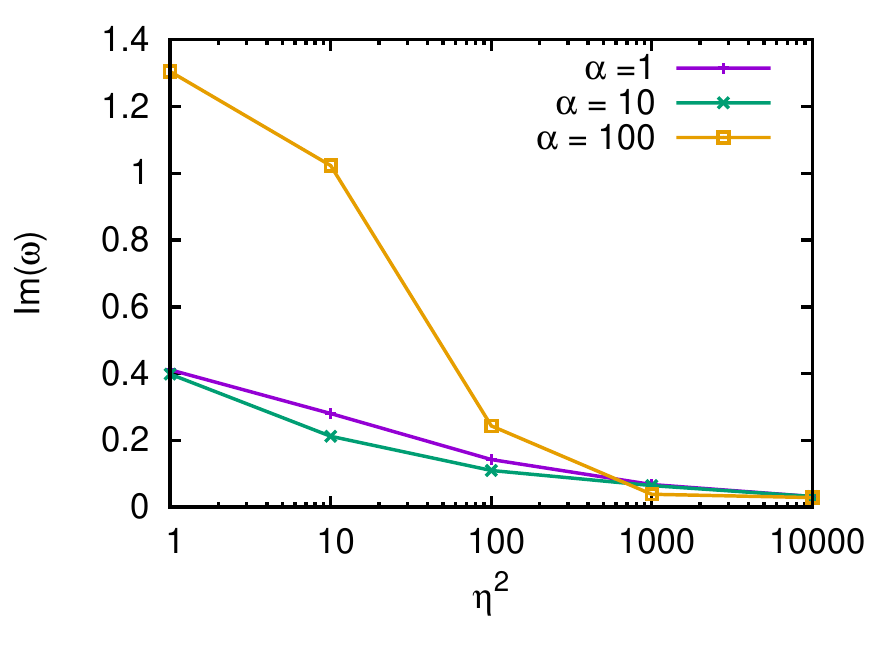} \\
\end{tabular}
\caption{Scalar quasinormal modes for a black hole, with a cloud of strings, in Einstein-Gauss-Bonnet gravity. The parameters values are $l=2$ and $n=0$, for $d=7$ (top) and $d=9$ (bottom). We vary both $\alpha$ and $\eta^2$.}
\label{EGBvariaEtaLarge}
\end{figure}

In figure (\ref{EGBvariaEtaLarge}) we plot the same set of data, but now we focus on higher values for the parameters $\alpha$ and $\eta^2$. The main effect of the cloud of strings is clear one more time: to decrease the value of the modes. The way how it happens, however, deserve some commentaries. First, for a value of the parameter $\eta^2$ high enough, the value for the modes becomes independent of the parameter $\alpha$. In other words, as the energy flux increases (as we consider a bigger density of strings), the constant of the Gauss-Bonnet term becomes unimportant to determine the value of the modes. In both graphs we can see that this happens when the parameter $\eta^2$ is about 10 times the value of the parameter $\alpha$. Also, the imaginary part of the quasinormal modes presents an unexpected behaviour, already noticed in figure (\ref{EGBvariaEta}). Usually, the imaginary part increases as the parameter $\alpha$ increases, but for some values of the parameter $\eta^2$, this behaviour changes, and the imaginary part of the modes becomes smaller for higher values of the parameter $\alpha$.  

\begin{figure}[]
\centering
\begin{tabular}{@{}cc@{}}
\includegraphics[scale=0.8]{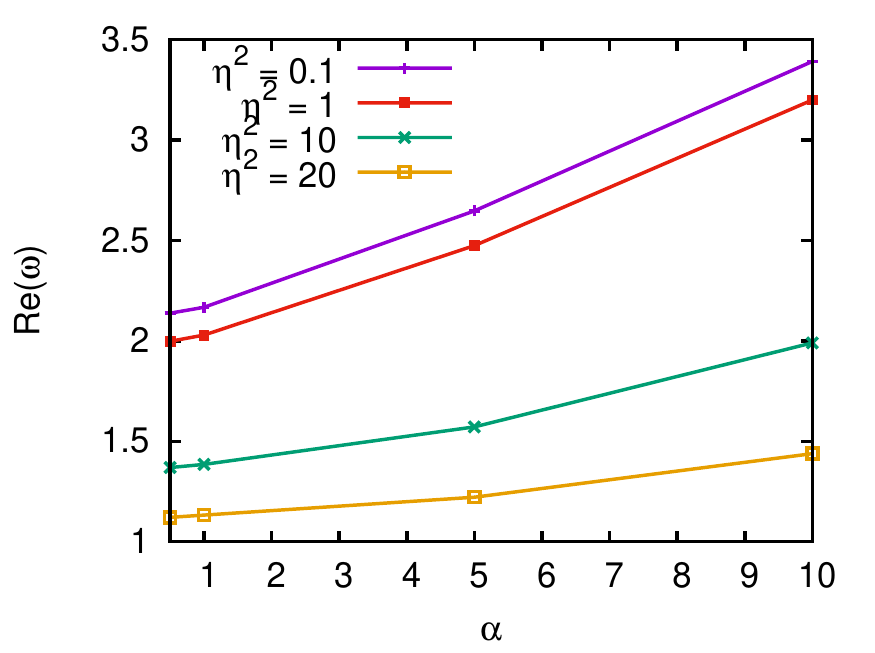} &
\includegraphics[scale=0.8]{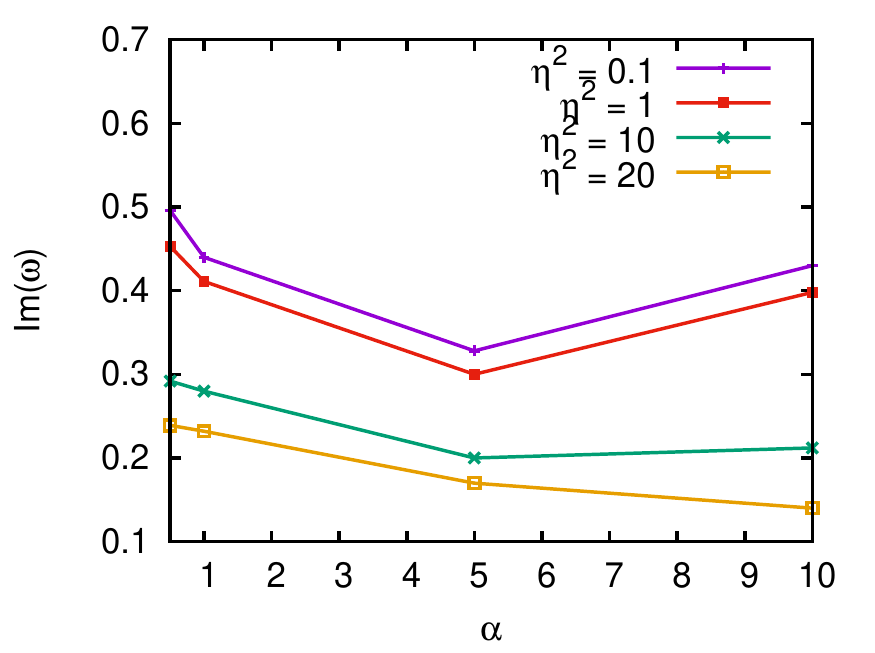} \\
\includegraphics[scale=0.8]{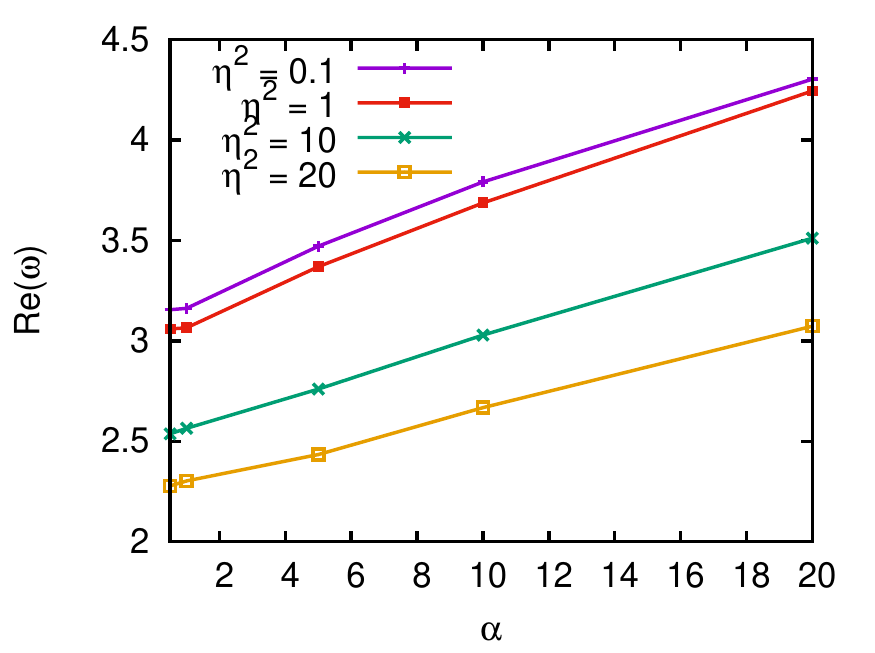} &
\includegraphics[scale=0.8]{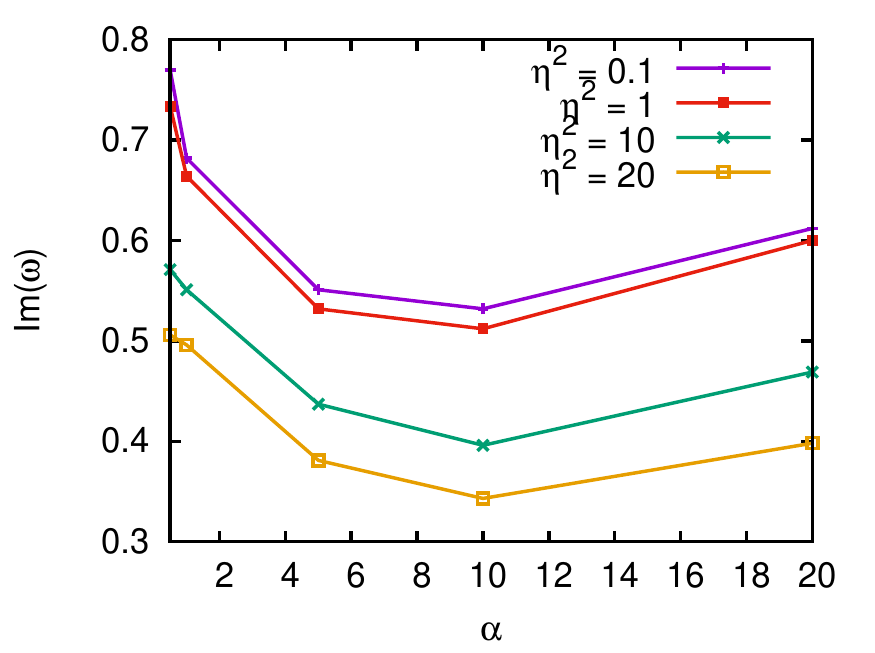} \\
\end{tabular}
\caption{Scalar quasinormal modes for a black hole, with a cloud of strings, in Einstein-Gauss-Bonnet gravity. The parameters values are $l=2$ and $n=0$, for $d=7$ (top) and $d=9$ (bottom). We vary both $\alpha$ and $\eta^2$.}
\label{fig:EGBvariaAlpha}
\end{figure}

Another way to explore this set of data is to fix the parameter $\eta^2$ and to vary the parameter $\alpha$. A set of four plots using this procedure is show in figure (\ref{fig:EGBvariaAlpha}), and the result is consistent with the ones presented in \cite{Konoplya:2004xx}, where the author studies the scalar quasinormal modes in Einstein-Gauss-Bonnet gravity. The real part increases with $\alpha$, but the imaginary part presents a more interesting behaviour. Untill some fixed value for $\alpha$, namely, $\alpha_0$, the value decreases, and then starts to increase. The presence of the string cloud can be noted in the decrease of the modes for a fixed $\alpha$, but also for changing the turning point $\alpha_0$. As the parameter $\eta^2$ increases, $\alpha_0$ also increases. In the next section we will study the same system under tensor perturbations.

\section{Tensor perturbations}

Tensor perturbations are obtained perturbing the metric in the gravitational field equations. This procedure for Einstein-Gauss-Bonnet theory in higher dimensions is not trivial. Fortunately, we can reduce this problem to a Schr\"odinger type equation (\ref{scalarEquation}), with the potential given by

\begin{equation}
V(r) = q(r) + \left(f(r)\frac{d}{dr}ln K(r)\right)^2 + f(r)\frac{d}{dr} \left(f(r) \frac{d}{dr}ln K(r) \right),
\end{equation}
with

\begin{equation}
K(r) = r^{\frac{d-4}{2}} \sqrt{r^2 + \alpha_2(d-4)\left[ (d-5)(1-f(r)) - r \frac{df(r)}{dr}\right]}
\end{equation}
and

\begin{equation}
q(r) = \left(\frac{f(2-\gamma)}{r^2}\right) \left( \frac{(1- \alpha_2 f'')r^2 + \alpha_2(d-5)[(d-6)(1-f)-2rf']}{r^2 + \alpha_2 (d-4)[(d-5)(1-f)-rf']} \right),
\end{equation}
where the radial dependence of the function $f(r)$ is implicit, and $\gamma = -l(l+d-3) + 2$. This procedure can be found in \cite{Dotti:2005sq}. 

\begin{figure}[]
\centering
\begin{tabular}{@{}cc@{}}
\includegraphics[scale=0.75]{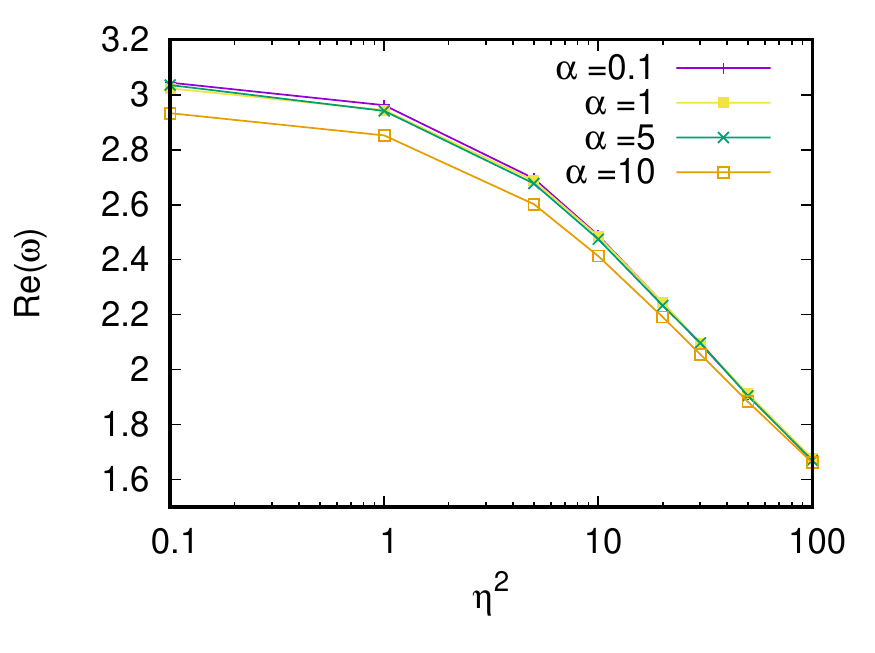} &
\includegraphics[scale=0.75]{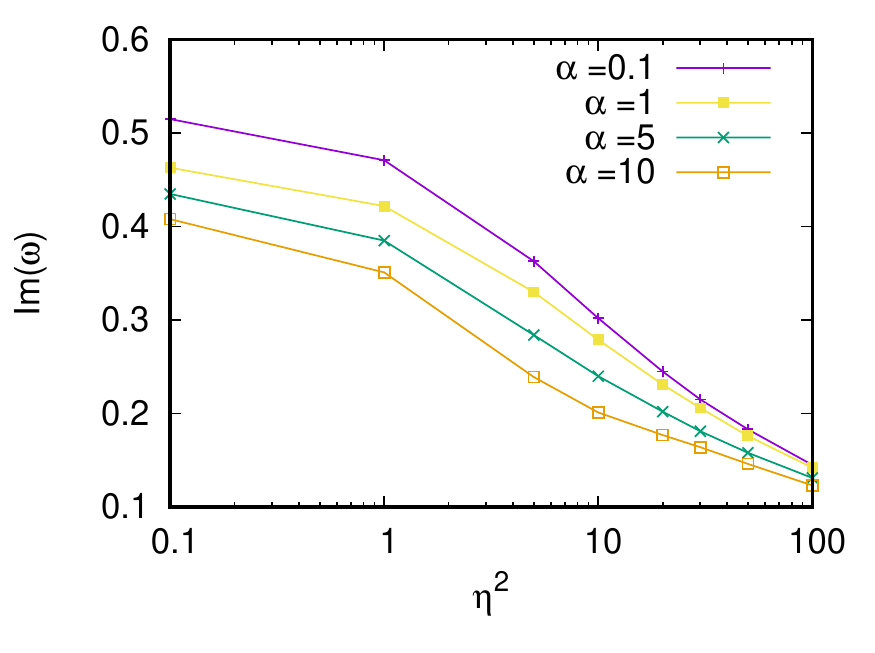} \\
\end{tabular}
\caption{Tensor quasinormal modes for a black hole, with a cloud of strings, in Einstein-Gauss-Bonnet gravity. The parameters values are $l=2$, $n=0$ for $d=7$.}
\label{TensorVariaEta}
\end{figure}

Some calculated values for the quasinormal modes, for some sets of parameters, are listed in table (\ref{tab:QNM_tensor}). It is more useful, however, to plot the data in a series of graphs. In figure (\ref{TensorVariaEta}) the real and imaginary values for the quasinormal modes are plotted for $l=2$, $n=0$ and $d=7$. The behaviour is the same as presented for the scalar modes. For $\eta^2 = 0$, the value for both the real and imaginary part decreases as the parameter $\alpha$ decreases, and both values decreases as we increase the value for the parameter $\eta^2$.

\begin{table}[]
\tbl{Tensor modes for $d=7$, $l=2$ and $n=0$.}
{
\begin{tabular}{p{0.2\textwidth}p{0.2\textwidth}p{0.2\textwidth}p{0.2\textwidth}}
\hline
 $\alpha$  & $\eta^2$  & Re($\omega$) & -Im($\omega$)  \\ \hline
 0.1  & 0.1  & 2.088 & 0.515  \\
 0.1  & 1  & 1.956 & 0.471  \\
 0.1  & 10  & 1.352 & 0.302  \\
 0.1  & 20  & 1.112 & 0.245  \\
 0.1  & 30  & 0.982 & 0.215  \\
 5.0  & 0.1  & 2.223 & 0.435  \\
 5.0  & 1  & 2.079 & 0.385  \\
 5.0  & 10  & 1.426 & 0.240  \\
 5.0  & 20  & 1.165 & 0.202  \\
 5.0  & 30  & 1.023 & 0.181  \\
 10.0  & 0.1  & 2.162 & 0.408  \\
 10.0  & 1  & 2.028 & 0.351  \\
 10.0  & 10  & 1.420 & 0.201  \\
 10.0  & 20  & 1.173 & 0.177  \\ 
 10.0  & 30  & 1.035 & 0.164  \\ 
\end{tabular}
\label{tab:QNM_tensor}
}
\end{table}

\begin{figure}[]
\centering
\begin{tabular}{@{}cc@{}}
\includegraphics[scale=0.75]{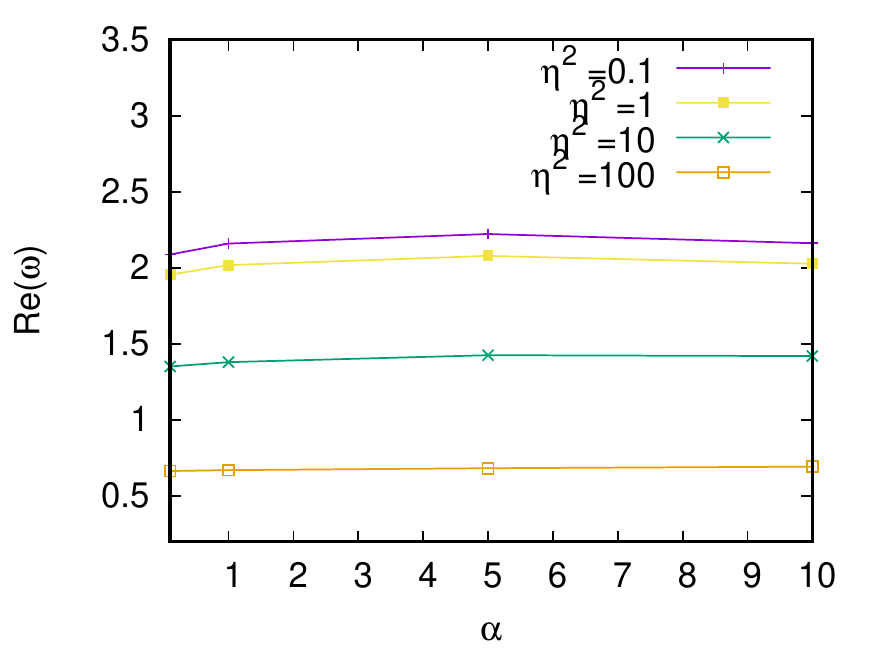} &
\includegraphics[scale=0.75]{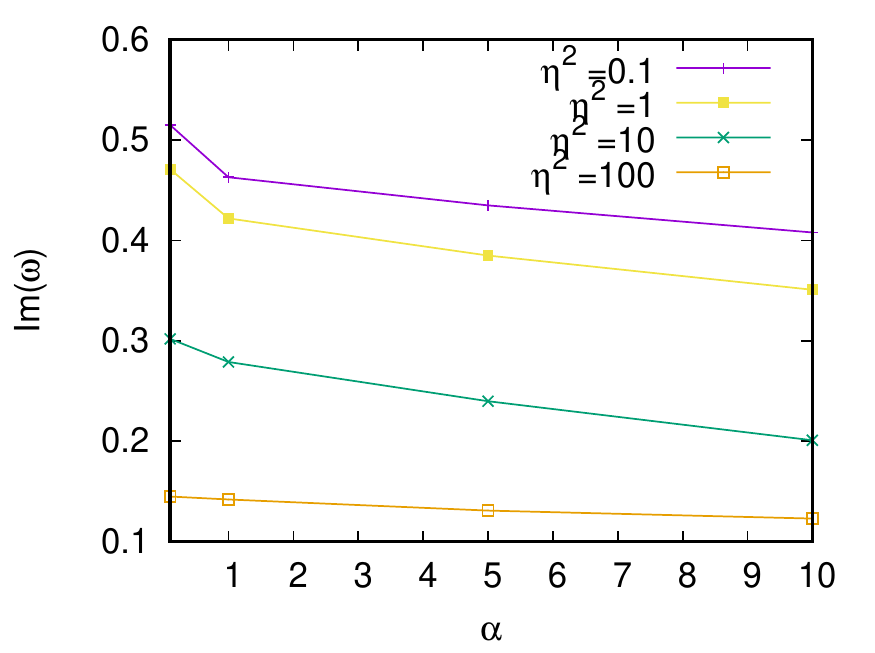} \\
\end{tabular}
\caption{Tensor quasinormal modes for a black hole, with a cloud of strings, in Einstein-Gauss-Bonnet gravity. The parameters values are $l=2$, $n=0$ for $d=7$.}
\label{TensorVariaAlpha}
\end{figure}

In figure (\ref{TensorVariaAlpha}) the parameter $\eta^2$ is fixed, and we vary the parameter $\alpha$. For the imaginary part, the values for the modes decreases as we increase the parameter $\alpha$, and the real  part does not show some definite behaviour. We had chosen to keep $\alpha$ not too greater than unity, since the system appears to be unstable for larger value of $\alpha$. We would like to call the attention to the fact that our results differs from the one obtained by \cite{Chakrabarti:2006ei}. The reason is the different parametrizations we used for the parameter $\alpha$. We believe we made the correct choice, since the tensor potential obtained in \cite{Dotti:2005sq} requires $\kappa = 1$, and not only $G=1$, as used in \cite{Chakrabarti:2006ei}.

We also consider as an important result the fact that the values for the scalar and tensor quasinormal modes approaches each other as the parameter $\eta^2$ increases. In the absence of the string cloud, there is not an isospectrality between the scalar and tensor modes, as noted in \cite{Chakrabarti:2006ei}. However, we can see that such isospectrality occur for large values of the parameter $\eta^2$, related with the string cloud.  

\section{Conclusions}

We have obtained the massless quasinormal modes for a black hole, with a cloud of strings, in Einstein-Gauss-Bonnet gravity for a scalar and a perturbed metric field. We have focused on the role played by the parameter $\eta^2$, related to the energy of the string cloud, and also by the parameter $\alpha$, which parametrize how the Einstein-Gauss-Bonnet gravity differs from general relativity. For the $\alpha \rightarrow 0$ and $\eta^2 = 0$, the spacetime solution is the Schwarzschild metric, and for $\eta^2 = 0$ and $\alpha \neq 0$, the solution is the Boulware-Deser metric.

We have found that, as $\eta^2$ grows, both the normal and imaginary parts of the modes decreases. This indicates that the frequency of emission is smaller when compared with the same system without the string cloud. It also indicates that the emission occurs for a longer period of time, since the damping (parametrized by the imaginary part of the QNMs) is smaller when the cloud of strings is present.

Another important result we can mention is related with the constant $\alpha_0$. As was already know \cite{Konoplya:2004xx,Chakrabarti:2006ei}, as we increase the parameter $\alpha$ starting from zero (the general relativity limit), the imaginary part of the QNMs (both scalar and tensor) initially decreases until it reaches the value $\alpha_0$, when it starts to increase. We denoted this \textit{turning point} by $\alpha_0$, and our analysis shows that $\alpha_0$ is a function of the energy of the cloud of strings, and increases as the energy increases.

One last result we should mention is when we compare the scalar and tensor QNMs. It was argued that these modes should be equal in vacuum \cite{Konoplya:2004xx}, but latter it was shown \cite{Chakrabarti:2006ei} that this was not the case, and that the scalar and tensor values of the QNMs become more and more different as $\alpha$ increases. We have shown that indeed the scalar and tensor QNMs are different, but as we increase the parameter $\eta^2$, these values approaches each other. Apparently, when $\eta^2 \gg \alpha$, the scalar and tensor modes acquire the same values. 

Unfortunately the obtained results cannot be compared with the observed gravitational wave signal \cite{Abbott:2016blz}. With the lack of some \textit{standard candle} for gravitational wave astronomy, it is impossible to perform some kind of \textit{reverse engineering} to extract parameters like $\alpha_2$ or $\eta$ (or, more in general, to claim validity of the Einstein-Gauss-Bonnet model). We hope that, in some future, high precision spectroscopy  near black holes will provide us enough data to constraint or allow such modified theories of gravity by its gravitational wave waveform.

\section*{Acknowledgments}
JPMG and VBB thanks Conselho Nacional de Desenvolvimento Científico e Tecnológico (CNPq) for partial financial support. Ines G. Salako thank IMSP for hospitality during the elaboration of this work.


\begin{thebibliography}{}
\bibitem{Copeland}
  E.~J.~Copeland and T.~W.~B.~Kibble,
  ``Cosmic Strings and Superstrings,''
  Proc.\ Roy.\ Soc.\ Lond.\ A {\bf 466}, 623 (2010)
  doi:10.1098/rspa.2009.0591
  [arXiv:0911.1345 [hep-th]].

\bibitem{Hindmarsh:1994re} 
  M.~B.~Hindmarsh and T.~W.~B.~Kibble,
  ``Cosmic strings,''
  Rept.\ Prog.\ Phys.\  {\bf 58}, 477 (1995)
  doi:10.1088/0034-4885/58/5/001
  [hep-ph/9411342].
  
  \bibitem{Letelier:1979ej} 
  P.~S.~Letelier,
  ``Clouds Of Strings In General Relativity,''
  Phys.\ Rev.\ D {\bf 20}, 1294 (1979).
  doi:10.1103/PhysRevD.20.1294
  
 \bibitem{Ghosh:2014dqa} 
  S.~G.~Ghosh and S.~D.~Maharaj,
  ``Cloud of strings for radiating black holes in Lovelock gravity,''
  Phys.\ Rev.\ D {\bf 89}, no. 8, 084027 (2014)
  doi:10.1103/PhysRevD.89.084027
  [arXiv:1409.7874 [gr-qc]].
  
\bibitem{Ghosh:2014pga} 
  S.~G.~Ghosh, U.~Papnoi and S.~D.~Maharaj,
  ``Cloud of strings in third order Lovelock gravity,''
  Phys.\ Rev.\ D {\bf 90}, no. 4, 044068 (2014)
  doi:10.1103/PhysRevD.90.044068
  [arXiv:1408.4611 [gr-qc]].
  
\bibitem{Mazharimousavi:2015sfo} 
  S.~H.~Mazharimousavi and M.~Halilsoy,
  ``Cloud of strings as source in $2+1$ -dimensional $f\left( R\right) =R^{n}$ gravity,''
  Eur.\ Phys.\ J.\ C {\bf 76}, no. 2, 95 (2016)
  doi:10.1140/epjc/s10052-016-3954-7
  [arXiv:1511.00603 [gr-qc]].     

\bibitem{Herscovich:2010vr} 
  E.~Herscovich and M.~G.~Richarte,
  ``Black holes in Einstein-Gauss-Bonnet gravity with a string cloud background,''
  Phys.\ Lett.\ B {\bf 689}, 192 (2010)
  doi:10.1016/j.physletb.2010.04.065
  [arXiv:1004.3754 [hep-th]].

\bibitem{Berti:2009kk} 
  E.~Berti, V.~Cardoso and A.~O.~Starinets,
  ``Quasinormal modes of black holes and black branes,''
  Class.\ Quant.\ Grav.\  {\bf 26}, 163001 (2009)
  doi:10.1088/0264-9381/26/16/163001
  [arXiv:0905.2975 [gr-qc]].

\bibitem{Abbott:2016blz} 
  B.~P.~Abbott {\it et al.} [LIGO Scientific and Virgo Collaborations],
  ``Observation of Gravitational Waves from a Binary Black Hole Merger,''
  Phys.\ Rev.\ Lett.\  {\bf 116}, no. 6, 061102 (2016)
  doi:10.1103/PhysRevLett.116.061102
  [arXiv:1602.03837 [gr-qc]].
  
\bibitem{Konoplya:2004xx} 
  R.~Konoplya,
  ``Quasinormal modes of the charged black hole in Gauss-Bonnet gravity,''
  Phys.\ Rev.\ D {\bf 71}, 024038 (2005)
  doi:10.1103/PhysRevD.71.024038
  [hep-th/0410057].
  
\bibitem{Chakrabarti:2006ei} 
  S.~K.~Chakrabarti,
  ``Quasinormal modes of tensor and vector type perturbation of Gauss Bonnet black hole using third order WKB approach,''
  Gen.\ Rel.\ Grav.\  {\bf 39}, 567 (2007)
  doi:10.1007/s10714-007-0404-8
  [hep-th/0603123].    
  
\bibitem{Cuyubamba:2016cug}
  M.~A.~Cuyubamba, R.~A.~Konoplya and A.~Zhidenko,
black holes in the de Sitter world,''
  Phys.\ Rev.\ D {\bf 93}, no. 10, 104053 (2016)
  doi:10.1103/PhysRevD.93.104053
  [arXiv:1604.03604 [gr-qc]].

\bibitem{Konoplya:2008ix}
  R.~A.~Konoplya and A.~Zhidenko,
  Phys.\ Rev.\ D {\bf 77}, 104004 (2008)
  doi:10.1103/PhysRevD.77.104004
  [arXiv:0802.0267 [hep-th]].

\bibitem{Abdalla:2005hu}
  E.~Abdalla, R.~A.~Konoplya and C.~Molina,
  Phys.\ Rev.\ D {\bf 72}, 084006 (2005)
  doi:10.1103/PhysRevD.72.084006
  [hep-th/0507100].  
  
\bibitem{Dotti:2004sh} 
  G.~Dotti and R.~J.~Gleiser,
  Class.\ Quant.\ Grav.\  {\bf 22}, L1 (2005)
  doi:10.1088/0264-9381/22/1/L01
  [gr-qc/0409005].  
  
\bibitem{Psaltis:2008bb} 
  D.~Psaltis,
  Living Rev.\ Rel.\  {\bf 11}, 9 (2008)
  doi:10.12942/lrr-2008-9
  [arXiv:0806.1531 [astro-ph]].  
  
\bibitem{Will} 
  Clifford M. Will ,
  Theory and Experiment in Gravitational Physics,
  Cambridge University Press, 1985

\bibitem{Calmet:2008tn} 
  X.~Calmet, S.~D.~H.~Hsu and D.~Reeb,
  Phys.\ Rev.\ D {\bf 77}, 125015 (2008)
  doi:10.1103/PhysRevD.77.125015
  [arXiv:0803.1836 [hep-th]].
  
\bibitem{Atkins:2012yn} 
  M.~Atkins and X.~Calmet,
  Phys.\ Rev.\ Lett.\  {\bf 110}, no. 5, 051301 (2013)
  doi:10.1103/PhysRevLett.110.051301
  [arXiv:1211.0281 [hep-ph]].
  
\bibitem{Xianyu:2013rya} 
  Z.~Z.~Xianyu, J.~Ren and H.~J.~He,
  Phys.\ Rev.\ D {\bf 88}, 096013 (2013)
  doi:10.1103/PhysRevD.88.096013
  [arXiv:1305.0251 [hep-ph]].
  
\bibitem{Onofrio:2010zz} 
  R.~Onofrio,
  Phys.\ Rev.\ D {\bf 82}, 065008 (2010)
  doi:10.1103/PhysRevD.82.065008
  [arXiv:1011.3064 [astro-ph.CO]].
  
\bibitem{Onofrio:2014txa} 
  R.~Onofrio and G.~A.~Wegner,
  Astrophys.\ J.\  {\bf 791}, 125 (2014)
  doi:10.1088/0004-637X/791/2/125
  [arXiv:1409.8546 [hep-ph]].
  
\bibitem{Wegner:2015aea} 
  G.~A.~Wegner and R.~Onofrio,
  Eur.\ Phys.\ J.\ C {\bf 75}, no. 7, 307 (2015)
  doi:10.1140/epjc/s10052-015-3523-5
  [arXiv:1503.01738 [astro-ph.IM]].          

\bibitem{Barriola:1989hx} 
  M.~Barriola and A.~Vilenkin,
  ``Gravitational Field of a Global Monopole,''
  Phys.\ Rev.\ Lett.\  {\bf 63}, 341 (1989).
  doi:10.1103/PhysRevLett.63.341  
  
\bibitem{Iyer:1986np} 
  S.~Iyer and C.~M.~Will,
  ``Black Hole Normal Modes: A {WKB} Approach. 1. Foundations and Application of a Higher Order {WKB} Analysis of Potential Barrier Scattering,''
  Phys.\ Rev.\ D {\bf 35}, 3621 (1987).
  doi:10.1103/PhysRevD.35.3621
  
\bibitem{Konoplya:2003ii} 
  R.~A.~Konoplya,
  ``Quasinormal behavior of the d-dimensional Schwarzschild black hole and higher order WKB approach,''
  Phys.\ Rev.\ D {\bf 68}, 024018 (2003)
  doi:10.1103/PhysRevD.68.024018
  [gr-qc/0303052].    

\bibitem{Dotti:2005sq} 
  G.~Dotti and R.~J.~Gleiser,
  ``Linear stability of Einstein-Gauss-Bonnet static spacetimes. Part I. Tensor perturbations,''
  Phys.\ Rev.\ D {\bf 72}, 044018 (2005)
  doi:10.1103/PhysRevD.72.044018
  [gr-qc/0503117].

\end{thebibliography}
\end{document}